\documentstyle[prd,aps]{revtex}
\begin{document}
\draft
\title{QUANTIZATION~OF~THE~DIRAC~FIBRE~:
A~NEW~WAY~IN~THE~PARTICLE~PHYSICS}
\author{S.~S.~Sannikov}
\address{Physico-Technical Institute\\
1 Academichna St., 310108 Kharkov, {\bf UKRAINE}}
\author{A.~A.~Stanislavsky}
\address{Institute of Radio Astronomy of
Ukrainian National Academy of Sciences\\
4 Chervonopraporna St., 310002 Kharkov, {\bf UKRAINE}\\
E-mail: stepkin@ira.kharkov.ua}
\date{\today}
\maketitle
\begin{abstract}
      The problem of ultraviolet divergences is analysed in
the quantum field theory. It was found that it has common roots
with the problem of cosmological singularity. In the context of
fibre bundles the second quantization method is represented as
a procedure of the quantization for vector bundle cross-section.
It is shown to be quite a different quantization way called as
a fibre quantization which leads to an idea on existence of the
non-standard dynamical system, i.e. the relativistic be-Hamiltonian
system. It takes place on supersmall distances and is well
described by the mathematical apparatus for the non-unitary quantum
scheme using a dual pair of topological vector spaces in terms of
the non-Hermitian form. The article contains the proof of the theorem
on radical changes in space and in matter structure taking place for
a very high density of matter: the phase transitions ``Lagrangian
field system (elementary particles) $\to$ relativistic
bi-Hamiltonian system (Feynman's partons)'' and ``continuum $\to$
discontinuum''.

      All required calculations in the framework of the proposed
theory are published in the Russian periodicals. The purpose of this
article is to replace the calculations by reasonings and concepts.
The present article begins the systematic exposition of principles
of the theory.
\end{abstract}
\pacs{11.}

\section{Introduction}
      Perhaps, many physicists agree with the opinion that the
elementary particle theory of today is in deep crisis.\footnote
   {A sign of the crisis is the contradiction between the urgent
necessity in a new physics related with supersmall distances or
particale structure and the continued utilization of old mathematical
instruments on the base of field concept (field is a function in
the space-time continuum), i.\ e.\ geometry, whereas the supersmall
distances belong to algebra rather than geometry.}
As will be shown below, it is caused by the early
application of the second quantization procedure.
To get it out of the state and to put an end to errors of
the last decades, it is necessary first to formulate
initial principles in their precise mathematical form.
Then using strict mathematical structures we shall show that
there is quite a different way in quantum field theory
by means of a new quantization procedure, the quantization of
Dirac fibre. As a result there appears an idea on existence of
a non-standard dynamical system in Nature. Proceeding from
the system, the consistent theory of elementary particles and
their interactions can be constructed.

\section{Initial statements}
      a) As is well known, since the thirties the quantum field
theory of particles confronted with the problem of ultraviolet
divergences. The problem being unsolved so far has blocked the
normal development of this theory. However, a cause of the
difficulty arose earlier. It turns out to be rooted in the second
quantization method using the distributions (for example, the
Dirac $\delta$-function).

      Up till now the physics uses the Newtonian concept of
space-time treated as a continuum or differential manifold $M_4$
at each point $X\in M_4$ of which there is a pair of vector
spaces --- the tangent space $T_X$ (with the basis
$\frac{\partial}{\partial X_\mu}$) and the cotangent
space $T^{*}_X$ (with the basis $dX_\mu$). $M_4$ is considered
to have the measure $d^{4}X$.

      Measure spaces are bad because they have sets of measure
zero\cite{1} (for example, isolated points in which the
point-like particles\footnote
     {It is important to note that with regard to the wave
properties (the quantization (\ref{eq1}) $\to$ (\ref{eq2}), see
below) the particles as though are smeared, and the degree of
divergences is reduced. Therefore, for the divergences be eliminated,
it is necessary to improve the quantum theory, rather than to
change the classical aspect of a particle approaching, for example,
to strings.}
can exist are the objects far from identical zero, but they are
alike by occuping the position in space-time). For such sets
will not quite lose in continuum, the notion of distributions
is introduced to mathematics. Namelly, the null sets are their
supports. From the point of view of functional analysis the
distrubutions are functionals or generalized functions.

      If ordinary functions (in particular, trial functions) form
not only a linear space (we may add them), but a ring too (we may
multiply them), distributions forming a linear space have not a
reasonable ring structure, i.\ e.\ in general, they do not admit the
multiplication operation. Really, for example, the product of two
$\delta$-functions, when their independent variables convergence
$\delta^2(X)$, is not a generalized function because their integration
with a trial function $f(X)$ yields (if $f(0)\not=0$) its infinite result:
$\int_{-\infty}^{+\infty}\delta^2(X)f(X)\,dX=\delta(0)f(0)=\infty$.

      Everything is all right till the generalized functions are
added only. But in the calculus of Feynman diagrams we must multiply
them. It reduces to the infinity which is called, differently, the
ultraviolet divergence.

      In view of this difficulty, it is first reasonable to
ask: Is it so necessary to applicate the second quantization
method (or anothers being equivalent to it) for particle fields?
Is it the only possible development for the quantum field
theory of particles?

      b) If the first (or space) quantization with a good empirical
foundation is the only possible mathematical foundation, then the
second (all the more third, fourth etc.) quantization can not
boast of this.

      The mathematical meaning of the first quantization is the
following. We pass from the phase space $PS^{2n}$ in the form of
the cotangent bundle $(M,T^*M)$, which is constructed over the
differential manifold $M$ (in a general case its dimension is
$n$), to the tangent bundle $(M,TM)$. Here $T^*=\bigcup_{X\in M}
T^*_X$, $T^*_X$ is the cotangent space at the point $X\in M$
with the basis $dX_j$ (in this case the momenta $P_j\,,\ j=
1,2,...,n$ is defined as $P_j\,dt=m\,dX_j$ where $t$ is the time,
$m$ is the mass of a particle), and $T=\bigcup_{X\in M}T_X$,
$T_X$ is the tangent space at the point $X\in M$ with the basis
$\frac{\partial}{\partial X_j}$ (in this case the momenta become
the operators $\hat P_j=-i\hbar\frac{\partial}{\partial X_j}$
where $\hbar$ is Planck's constant).

      To transform $(M,T^*M)$ and $(M,TM)$ in a canonical system,
in the first (classical) case the Lie structure is known to be given by
Poisson brackets $\left\{f,g\right\}=\Sigma^n_{j=1}(\frac{\partial f}
{\partial X_j}\frac{\partial g}{\partial P_j}-\frac{\partial f}
{\partial P_j}\frac{\partial g}{\partial X_j})$
where $f$, $g$ are functions in $(M,T^*M)$ as
\begin{equation}
\left\{X_j,P_k\right\}=\delta_{jk},\quad\left\{X_j,X_k\right\}=
\left\{P_j,P_k\right\}=0,\label{eq1}
\end{equation}
and in the second (quantum) case the Lie brackets (commutators) are used
\begin{equation}
[\hat X_j,\hat P_k]=i\hbar\left\{X_j,P_k\right\}=i\hbar\delta_{jk},
\quad[\hat X_j,\hat X_k]=[\hat P_j,\hat P_k]=0.\label{eq2}
\end{equation}
Relations (\ref{eq1}) and (\ref{eq2}) define the Heisenberg
algebra $h_{2n}$.

      Being accompanied by the mapping (\ref{eq1}) $\to$ (\ref{eq2})
the mapping $(M,T^*M)\to(M,TM)$ is uniquely, universal and the
only possible mapping, i.\ e.\ posseesses the functor properties.
Thus two and only two physics theories --- classical $(M,T^*M)$
and quantum $(M,TM)$ --- can be related with the continuum $M$.
Then, the dynamical system is defined as a canonical system with a
given dynamical group. In the case of small oscillations (plaing a
particularly important role in the microcosm physics) the dynamical group
is the group $Sp(n)$ of automorphisms for the Heisenberg algebra $h_{2n}$
\cite{2}. In our opinion it should be error to consider, in connection
with the microcosm, the nonlinear (curved) phase spaces and so-called
geometric quantization.

      We can now say that the first quantization is a functor in the
category of differential manifolds from $(M,T^*M)$ to $(M,TM)$.
The second quantization is mathematically empty (see below) as
opposed to the first quantization.

      c) Next, both classical and quantum theories consider the
enveloping algebras $U[(M,T^*M)]$ for $(M,T^*M)$ and $U[(M,TM)]$ for
$(M,TM)$ as over Heisenberg algebras $h_{2n}$. Note that it is
the pure algebraic structure. In this case the dynamical variables of
one system or another belong to these algebras or to their certain
topological closures.

      The quantum theory deals with a representation of the associative
algebra $U[(M,TM)]$ ($U$ is the operation taking an enveloping algebra)
in the topological vector space $\cal F$. The space is constructed
as follows. The maximal commutative subalgebra --- the Lagrange plane
--- is considered in the Heisenberg algebra $(M,TM)$. Usually $M$
itself takes its role \cite{4}. The enveloping algebra $U[M]$ is
constructed for $M$ as a maximal commutative subalgebra in
$U[(M,TM)]$. $U[M]$ is taken to be a compact set in $\cal F$, and
$\cal F$ itself is obtained from $U[M]$ by means of the
topological closure along the topology $\tau$:
${\cal F}=\overline{U[M]}^\tau$. As a rule, the Hilbert topology and
the Hilbert space $H$ are considered. But the extended Hilbert spaces
${\cal F}\subset H\subset {\cal F}'$ are also considered
where $\cal F$ and ${\cal F}'$ are the spaces of trial and generalized
functions (distributions in $M$ belong to ${\cal F}'$, but if
$H$ is a ring, see \cite{1}, its expansion ${\cal F}'$ is no
longer a ring).

      The elemements $\psi(X)$ ($X\in M$) of the space $H$ are called
the wave functions or state vectors of the system $(M,TM)$. The
Hermitian definite form is a scalar product in $H$
\begin{equation}
(\psi,\psi')=\int \psi^*(X)\psi'(X)\,dX \label{eq3}
\end{equation}
where $dX$ is the measure for $M$. The representation $U[(M,TM)]$
in $H$ is called the Schr$\rm\ddot o$dinger one. Note that
in this representation the state vectors of the system are
functions in $M$. The Hermitian definite form (\ref{eq3}) plays
the most important role in the Heisenberg-Schr$\rm\ddot o$dinger
quantum theory: it corresponds to the unitarity axiom defining
the unitary nature of used symmetry group representations
of the space $M$.

      d) In the case of elementary particles with spin
a wholly different mathematical structure, namely the
geometric structure of vector bundle over $M$, is used
to construct the space of state vectors. The vector
bundle is the trio $E=(M,S,\bar L)$ where $M$ is the base
(a differential manifold with a symmetry group $L$), the fibre
$S$ is the vector space, and $\bar L$ is the structure group
(covering for $L$) of the fibre $S$. In the particle theory
$M=\bf A_{3,1}$ is the affine space-time with the inhomogeneous
Lorentz group (or Poincar$\rm\acute e$ group) as a symmetry
group, and in the case of the most fundamental particles ---
fermions\footnote
          {It is important to note that the Dirac structure
of extracting of the square root of $TM$ leads to a notion
of the most fundamental Dirac (or spinor) fibre as to opening the
spin variables of particles (they do not reduce to the space
variables of $(T,TM)$ that is of great importance)\cite{4}. As a
result the mapping of the vector space $TM$ (or $M$) is the Clifford
algebra $C$ for the Dirac matrices $\gamma_\mu$: $TM\to C$. Then
the vectors, specifically current $j_\mu$, are constucted by the
formula $\bar\psi\gamma_\mu\psi$ ($\psi$,$\bar \psi$ are the elements
of the Dirac fibre) describing the quantization of the current
$j_\mu$.}
--- $S$ is the space of the Dirac bispinors $S^{(*)}_8\ni\psi_\alpha,
\bar \psi_\alpha$ (where $\bar \psi=\psi^*\gamma_4$ is the Dirac
adjoint bispinor to $\psi$, and $*$ is the involution connected
with this adjoint or simply the complex adjoint operation).

       The wave functions or particle fields $\psi(X)$, $\bar \psi(X)$,
satisfying the differential equations, are the cross-sections of the
bundle $E=(M,S,\bar L)$. Thus in the context of fibre bundles the field
$\psi(X)$ in $M$ is the cross-section of the conformable vector bundle
(all definitions used here can be found, for example, in \cite{5}),
and the Hilbert space $H$ is a space of bundle cross-sections.

      e) Usually, the main development of the theory is associated
with the quantization of the space ${\cal F}'$ , i.\ e.\ the quantum
postulate (second quantization method) is employed to bundle
cross-sections. In our opinion the use of this procedure is early
in the position.

      What reasons proves the need of the second quantization?
Only by that if the field $\psi(X)$ satisfies the Klein-Gordon
equation $(\Box-m^2)\psi(X)=0$ then its harmonics $\psi({\bf P},t)$,
in terms by which $\psi ({\bf X},t)$ is expressed as
\begin{displaymath}
\psi(X)=\frac {1}{2\pi^{3/2}}\int(e^{i{\bf P X}}
\stackrel {(+)}{\psi}({\bf P},t)+e^{-i{\bf P X}}
\stackrel {(-)}{\psi}({\bf P},t))\,d^3 {\bf P}\,,
\end{displaymath}
satisfy to the oscillator-like equation\cite{6}:
$\ddot\psi({\bf P},t)+P^2_\circ \psi({\bf P},t)=0$
where $P_\circ=\sqrt{{\bf P}^2+m^2}$. Hence,
the same commutation relations, which the first quantization
put to variables of oscillator, could be applicable to $\stackrel
{(\pm)}{\psi}({\bf P},t)$. But it is not, generally speaking,
true. The form of commutation relations for fields must
depend on their spin and statistics and does not
always reduce to the commutation relations for oscillator. In this case,
the analogy with oscillator is very superficial, and the quantization
procedure of fields does not follow uniquelly from it. And if the field
$\psi(X)$ satisfies not the Klein-Gordon equation but, say, the Bopp
equation $(\Box-M^2)(\Box-\mu^2)\psi(X)=0$, generally, there is no
analogy with an oscillator.

      As we already known, the second quantization method is bad
because of using distributions in commutation relations, and it yields
ultraviolet divergences. We see that in the context of fibre bundles
the second quantization method presents the quantization procedure of
bundle cross-sections, i.\ e.\ the mapping $\psi(X)\to\hat\psi(X)$
which we write as ${\cal F}'\to \hat {\cal F}'$ (elements of the set
$\hat {\cal F}'$ are local field operators $\hat\psi(X)$).
At the same time it is required for $\hat {\cal F}'$ to be as a ring
(algebra). It should be noted that since ${\cal F}'$ being a set of
distributions (fields) in $M_4$ is not a ring, $\hat {\cal F}'$
is not one too. And a smoothing of the form
$\hat \psi(f)=\int \hat\psi(X) f(X)\,dX$ is no effective
because the integration of bad (non-bounded) operators such as
$\hat \psi(X)$ can make the operator $\hat\psi(f)$ more
worse.\footnote{Conversely, the derivation
$F(\frac{\partial}{\partial X})\hat\psi(X)$, see \cite{8},
can improve the operator. It is surprising that in
the axiomatic approach the so-called smoothing operators are
used everywhere see, e.\ g.\ \cite{7}.}

      The mapping ${\cal F}'\to\hat {\cal F}'$ is very
contradictory. If the unitarity axiom holds true then
the local field operator $\hat\psi(X)$ does not take place at
all as a basis of the theory of quantized fields(Wightman's
theorem \cite{7}). It means that any quantization procedures
of infinite-number-of-degrees-of-freedom system is
inadmissible in principle. Nevertheles, for example, in
the case of the most fundamental Dirac fields $\psi(X)$, the
equal-time commutation relations are written in the form
\begin{equation}
\left\{\hat\psi({\bf X},t),\hat{\bar\psi}({\bf X'},t)\right\}=
\gamma_4\delta^3({\bf X}-{\bf X'}).\label{eq4}
\end{equation}
Such relations define the continual Clifford algebra
(in the case of bosons it will be the continual Heisenberg algebra).
At ${\bf X}={\bf X'}$ (i.\ e.\ when $\psi$ ¨ $\bar \psi$ are taken
from the same fibre) we have $\{\psi,\bar \psi\}=\gamma_4
\delta^3(0)=\infty$.

      Heisenberg was a first who understood \cite{9} that the cause of
all ultraviolet divergences was the $\delta$-function on the right-hand
side of (\ref{eq4}).

      We should put an end to ultraviolet divergences once and for all
if, following Heisenberg \cite{9}, we rejected the $\delta$-function
on the right-hand side of (\ref{eq4}). Finally, we should arrive
at such relations
\begin{equation}
\left\{\hat\psi({\bf X},t),\hat{\bar\psi}({\bf X'},t)\right\}=0
\label{eq5}
\end{equation}
which are valid for ${\bf X}={\bf X'}$ too,
i.\ e.\ at each isolated point of the space $M_4$. Notice that
relations (\ref{eq5}) for fixed $X$ define the so-called
finite dimensional Grassmann algebra. Regretfully (rather
fortunately), as Pauli has noted, relations (\ref{eq5})
contradict the differential equations of the first order
(it does not matter linear or nonlinear) for the field
$\hat\psi(X)$. The real way to overcome this contradiction is
only one: to give up any of differential equations in $M_4$.
But then the space in which we may not differentiate and integrate
of course is not now the Newtonian space. Next, we shall show
that the algebraic contraction (\ref{eq4})$\to$(\ref{eq5}) is
associated with the rejection from the Newtonian concept
of space-time as a differential manifold
in favour of the Riemannian idea on the completely
spatial non-connection. The contraction occurs if and only if
the space-time as a continuum is transformed into discontinuum,
i.\ e.\ if there takes place the phase transition ``continuum
$\to$ discontinuum'' which must come under certain
extremal physical conditions, see section \ref{kd}.

\section{Phase transition ``continuum $\to$ discontinuum''}\label{kd}
      Since in obvious way the problem of ultraviolet divergences
is related with small distances now we pay attention mainly to them.
Apparently, the supersmall distances must obey their particular physics
which is incorrectly described by the local field theory. But what a
specific character have supersmall distances compared, for example,
with atomic or nuclear distances (including also distances related
to dimensions of elementary particles)?

      To answer this question we consider the state of matter
characterized by its supercompact packing of elementary particles
(in natural conditions the situation arises for collapsing
Universe, i.\ e.\ in the neighbourhood of the singular point
$R_{\rm cr}=0$, when its density of mass begins to exceed the density
of nuclear matter $\rho_{\rm nuc}\sim$10$^{15}$ g/cm$^3$ by many
orders and, as calculations have been shown in \cite{10},
reaches the value $\rho_{\rm cr}\sim$ 10$^{30}$ g/cm$^3$,
whereas in laboratory conditions this is also attained for
collisions of particles with very high energies
$\varepsilon\sim10^5$ GeV, \cite{10}).

      This state is characterized by the space between particles
becoming less and less (in the critical point $R_{\rm cr}=0$
it vanishes at all, thus the main sign of the Hamiltonian or
Lagrangian property of the system disappears completely: the
Lagrangian plane $M$ concentrates in one point).

      In such extremal conditions (since there are no free seats)
the space-time translations defined on fields of particles
(i.\ e.\ in quantum theory) by the formula $\psi(X)\to\psi(X+a)=
e^{a\frac{\partial}{\partial X}}\psi(X)$ become impossible
transformations, whatever $a$. This means that the operator
$e^{a\frac{\partial}{\partial X}}$ does not exist exactly
as generators of these transformations
$\frac{\partial}{\partial {X_\mu}}$. Since
$\frac{\partial}{\partial {X_\mu}}$ form a basis in the
tangent space at point $X\in M$ from this it follows that
in the situation under consideration the tangent spaces (and
hence the cotangent spaces of vectors $dX_\mu$) do not exist.
All this means that the space $M$ loses its former structure (of
differential manifold) postulated by Newton, disintegrates on its
isolated points, becomes a completely non-connected set of its
points (in the case it continues to consist as consisted of
infinite uncountable numbers of points), i.\ e.\ continuum is
transformed into discontinuum (the specific character of supersmall
distances lies in this).

      In this conditions the Newtonian concept of space as a continuum
(as a differential manifold) does not work any more and must be
replaced by the Riemannian discrete concept (more precisely, of
completely non-connection) of space in a little (i.\ e.\ for supersmall
distances or superhigh energies).

      Mathematically, the transition ``continuum $\to$ discontinuum''
signifies that we consider the greatest discrete topology for $M$
which is defined as the neighbourhood of any point does not contain
the others.

      In the theory of particles and fields the space-time continuum
is an affine space $\bf A_{3,1}=( A_{3,1},R_{3,1} )$ where $\bf R_{3,1}$
is the vector Minkowski space (associated with $\bf A_{3,1}$) with
the Poincar$\rm\acute e$ group ${\cal P}=L\,\times\!)\,T_{3,1}$ as
a symmetry group where $L$ is the inhomogeneous Lorentz group
(the symmetry group of the space $\bf R_{3,1}$), and $T_{3,1}$ is
the group of translations in $\bf A_{3,1}$. When $\bf A_{3,1}$
disintegrates on its isolated points the group $T_{3,1}$
collapses only, the symmetry comes lower from $\cal P$ to $L$.
      By analogy we may say that for the superhigh
density of energy ``the mathematical fluid'' --- continuum ---
is coming to the boil changes into ``the gas'' --- discontinuum.

      When the base $M$ becomes the completely non-connected space,
the vector bundle $E=(M,S,\bar L)$ disintegrates in its isolated
fragments --- fibres $S_X$ given at each isolated point $X\in M$.
Here it is important to note that a point $X$ with its fibre
$S_X$ is more fundamental (and profound) object than the point
$X$ without one. If $M$ may be identified by an empty space
then $E$ should be a space with matter.

      Let us see now in what are transformed the commutation relations
(\ref{eq4}) for the transition ``continuum $\to$ discontinuum''(we
shall call it by the phase transition).
In new conditions the Dirac $\delta$-function
$\delta^3({\bf X}-{\bf X'})$ having the dimension cm$^{-3}$ passes
to the dimensionless Kronecker symbol $\delta_{\bf X,X'}$.
Since the separated points have the null measure and size, both
continuum and discontinuum are not characterized by any fundamental
lenght. Outgoing from physical reasons, we put
\begin{equation}
\left\{\hat\psi({\bf X},t),\hat{\bar\psi}({\bf X'},t)\right\}=
\gamma_4\left(mc\over \hbar\right)^3\delta_{\bf X,X'}
\label{eq6}
\end{equation}
where $m$ is the mass of the particle described by the field
$\psi(X)$. But since the mass of either particle for superhigh
energies may be ignored (the modulus square of field on the left
of (\ref{eq6}) is much more than the right side of this formula)
and because in the completely non-connected space the
matter does not exist as a particles we have nevertheles
another relations, namely
\begin{displaymath}
\left\{\hat\psi({\bf X},t),\hat{\bar\psi}({\bf X'},t)\right\}=0.
\end{displaymath}
Thus we have arrived at the Grassmann algebra, so we may say
that the phase transition ``continuum $\to$ discontinuum'' is
accompanied by the contraction of algebra (\ref{eq4}) to
algebra (\ref{eq5}).

      Hence, in the case when the space-time $M$ becomes a
discontinuum the Dirac fibre $S^{(*)}_8$ must be considered for
the Grassmann algebra $G$: $S^{(*)}_8=S^{(*)}_8(G)$.

      The phase transition ``continuun $\to$ discontinuum'' with
certain care can be called the quantization of space. At the
same time it should be noted that if the quantum nature of matter
shows itself enough early, for moderate energies $\sim$ 1 eV $\div$
10$^6$ eV (the quantum ladder has several steps: molecules, atoms,
nuclei, elementary particles) the quantum nature of space begins
to show itself much later, only for energies $\sim 10^{14}$ eV.
In the quantized space --- discontinuum --- there is no both the
measure $dX$ and Hermitian form (\ref{eq3}). Thus in this
case the unitarity axiom loses validity, and now we must use the
non-unitary symmetry group representations of the space $M$,
which were discovered in \cite{10} for the physically important
case of spaces $\bf R_3$, $\bf R_{3,1}$ (i.\ e.\ for groups
$SO(3)$ and $SO(3,1)$).

\section[Phase transition ``LFS $\to$ RBS'']{Phase transition ``Lagrangian
field system $\to$ relativistic bi-Hamiltonian system''}
      As shown in \cite{10}, the Dirac-Grassmann fibre $S^{(*)}_8(G)$
has a complex internal, inherently dynamical, structure.
Now we reconstruct the internal evidence.

      The dymanical structure of the fibre $S^{(*)}_8(G)$ is
established by the splitting of skew-symmetric (symplectic)
quadratic form for $S^{(*)}_8(G)$ in linear forms. It is interesting
to note that this structure adjoins the number theory in the
spirit of Galois's and Kummer's investigations corresponding to
search of the most fundamental numbers which control the Universe.
On the way to the realization of this idea the Grassmann numbers
were discovered. Grassmann was right in regard to
the applicability field of his numbers, this is the microcosm.
However, other numbers control the submicrocosm.

      A sympletic form on $S^{(*)}_8(G)$ is written as
$[\chi,\chi]=\chi E\chi=-2\bar\psi\psi\not\equiv 0$ where
$\chi={\phi\choose \bar\phi}$ is the 8-spinor,
$E=\pmatrix{\bf 0&\bf1\cr \bf-1&\bf0}$ and $\bf 1$ ($\bf 0$) is
the unit (null) 4å4-matrix. The form $[\chi,\chi]$ as a polinomial
of second degree for $\chi$ can be factorized into linear forms,
by writting $[\chi,\chi]=\hat\chi^2$ where $\hat\chi=\sqrt2
A_\alpha\chi_\alpha$ is the linear form. The coefficients
$A={\phi\choose\bar\phi}$ of the linear form should take values
from Heisenberg algebra. Since if $X\in G$ then, evidently, the
relations $A_\alpha A_\beta-A_\beta A_\alpha=E_{\alpha\beta}$
must be valid as commutation relations
\begin{equation}
[\phi_\alpha,\bar\phi_\beta]=\delta_{\alpha\beta},\quad
[\phi_\alpha,\phi_\beta]=[\bar\phi_\alpha,\bar\phi_\beta]=0
\label{eq7}
\end{equation}
which define the Heisenberg algebra $h^{(*)}_8$ with the
involution $*$ (one among real forms of the complex algebra
$h_8(G)$ which we have called the Dirac form). The mapping
$\psi_\alpha\to\phi_\alpha$, $\bar\psi_\alpha\to\bar\phi_\alpha$
in writting $S^{(*)}_8\to h^{(*)}_8$ is called the quantization of
Dirac fibre. Thus, we arrive at the Heisenberg algebra as well as to
the canonical system for which its canonical variables are the generators
$\phi_\alpha$, $\bar\phi_\alpha$ of the algebra $h^{(*)}_8$.

      Notice that in this case the Heisenberg algebra plays the same
role as algebraically closed rings in Golois's theory and
represents an analogy of Clifford algebra which arises by
means of the Dirac operation employed to the vector space for the
ring of usual Euclidean numbers. At the same time the elements of the
representation space of the algebra $h^{(*)}_8$ (we called the
semispinors, see below) play the role of ideal Kummer's numbers from
which are constructed the tensorial values: spinors, scalars, vectors
and so on.

      Taking into account that canonical variables $\phi$, $\bar \phi$
exist inside a spinor fibre which is considered at an isolated
space-time point (i.\ e.\ for supersmall distances where
space-time is discontinuum), and hence they should be placed into
submicrocosm, we formulate the dynamical principle: the dynamical
group of our system is the group of automorphisms for the algebra
$h^{(*)}_8$, which is denoted by
$Sp^{(*)}(4,\bf C)$. The generators of the dynamical group
are the every possible bilinear forms of canonical variables:
$\bar\phi_\alpha\phi_\beta,\phi_\alpha\phi_\beta,
\bar\phi_\alpha\bar\phi_\beta$ (quadratic Hamiltonians) forming
the semisimple Lie algebra which is isomorphic to the Cartan algebra
$sp^{(*)}(4,\bf C)$. To study the properties of the dynamical group
and its linear representation we may say much about the properties of
the dynamical system itself. We pay attention to the most important
ones of them.

        The real dynamical variables which we can write by means of
sixty Dirac matrices $\gamma_N$ as $\bar\phi\gamma_N\phi$ play an
especially important role. They form the Lie algebra which is isomorphic
to the algebra $u(2,2)$ (this isomorphism is given by the mapping
$\gamma_N\to\bar\phi\gamma_N\phi$). Among them there is the pair of
4-vectors
$\Gamma_\mu=i\bar\phi\gamma_\mu P_+\phi$ ,
$\dot\Gamma_\mu=-i\bar\phi\gamma_\mu P_-\phi$ where
$P_\pm=\frac{1}{2}(1\pm\gamma_5)$ having all properties of
4-momenta (excepting their dimension;
$\phi$, $\bar\phi$ are the dimensionless values)
\begin{displaymath}
[\Gamma_\mu,\Gamma_\nu]=[\dot\Gamma_\mu,\dot\Gamma_\nu]=0,
\end{displaymath}
but non-commuting with each other:
\begin{displaymath}
[\Gamma_\mu,\dot\Gamma_\nu]=2i I_{\mu\nu}+\delta_{\mu\nu}B
\end{displaymath}
where $I_{\mu\nu}=\bar\phi\Sigma_{\mu\nu}\phi$ (
$\Sigma_{\mu\nu}=\frac{1}{4i}[\gamma_\mu,\gamma_\nu]$ ),
and $B=\bar\phi\gamma_5\phi$. The values $I_{\mu\nu}$, $B$
and $A=\bar\phi\phi$ form the Lie algebra which is isomorphic to
the Lie algebra $gl(2,\bf C)$. We define the 4-momenta
$p_\mu,\dot p_\mu$ as $p_\mu=k\hbar\Gamma_\mu$,
$\dot p_\mu=k\hbar\dot\Gamma_\mu$ where $\hbar$ is Planck's constant,
and $k$ is the third (after $c$ and $\hbar$) fundamental constant having
the dimension  cm$^{-1}$. It follows from the definition of
$\Gamma_\mu$ and $\dot\Gamma_\mu$ that $\Gamma_\mu$ and
$\dot \Gamma_\mu$ are the isotropic 4-vectors, i.\ e.\
$p^2_\mu=\dot p^2_\mu=0$.

      Thus the group $U(2,2)$ consists of two different
Poincar$\rm\acute e$ subgroups ${\cal P}=GL(2,{\bf C})\,\times\!)\,T_{3,1}$
and $\dot {\cal P}=GL(2,{\bf C})\,\times\!)\,\dot T_{3,1}$ crossing along
the inhomogeneous Lorentz group $GL(2,\bf C)$. In the dynamical group
the operators ($I_{\mu\nu}$,$A,B$), $p_\mu$ and $\dot p_\mu$ are
the generators for $GL(2,\bf C)$, $T_{3,1}$ and $\dot T_{3,1}$.

      The systems the dynamics of which is not described by one
4-momentum (as in the case of Hamiltonian systems), but a pair of those
non-commuting with each other are called as relativistic be-Hamiltonian
ones. Our analysis of the Dirac fibre structure allows to say that
in extremal conditions  when the elementary particles are in the highly
compressed state and when all space is transformed from continuum into
discontinuum (see section \ref{kd}) the matter changes radically:
it is transformed from the Lagrangian system into the bi-Hamiltonian one.

      The dynamics of bi-Hamiltonian matter is written by a pair of
Hamiltonian flows non-commuting one with the other. In the Heisenberg
picture (in the enveloping algebra $U[h^{(*)}_8]$) these flows are
written by equations
\begin{equation}
\cases{-i\frac{\partial}{\partial x_\mu}F=[p_\mu,F]\cr
-i\frac{\partial}{\partial \dot x_\mu}F=[\dot p_\mu,F]\cr}
\label{eq8}
\end{equation}
where $F\in U[h^{(*)}_8]$, $x$ and $\dot x$ are the coordinates
on groups $T_{3,1}$ and $\dot T_{3,1}$ (resulting from
the consideration of the automorphisms $e^{ipx}Fe^{-ipx}$ and
$e^{i\dot p\dot x}Fe^{-i\dot p\dot x}$). It is easy to see
that system (\ref{eq8}) is not integrable in terms of
$$\left(\frac{\partial}{\partial x_\mu}\frac{\partial}
{\partial\dot x_\nu}-\frac{\partial}{\partial\dot x_\nu}
\frac{\partial}{\partial x_\mu}\right)\,F\,\neq\,0\,,$$
so that the operator $F(x,\dot x)$ does not exist as
a function on the manifold $U(2,2)/GL(2,\bf C)$.

      The same result is obtained in the Schr$\rm\ddot o$dinger
picture when it is considered in the unitary scheme using the
self-adjoint space $H$:
the system
\begin{equation}
\cases{-i\frac{\partial}{\partial x_\mu}f=p_\mu f\cr
-i\frac{\partial}{\partial \dot x_\mu}f=\dot p_\mu f\cr}
\label{eq9}
\end{equation}
where $f\in H$ is not integrable; the vector $f(x,\dot x)$ does not
exist as a function of two variables $x$ and $\dot x$. Such functions
are indispensable in connection with reconstruction of the
manifold $M=\bf A_{3,1}$.

      As is shown above, the unitary scheme postulated by von
Neumann at his time in connection with needs of the
Heisenberg-Schr$\rm\ddot o$dinger quantum theory does not work in
our case. The cause lies in the motion equations being not integrated
on a semisimple Lie group. On a solvable group the equations are
integrated only. But a semisimple noncompact group as
$Sp^{(*)}(4,\bf C)$ has always solvable groups. As is well known, these
are Borel subgroups associated with the Gaussian decomposition $N_-HN_+$
for group. For this reason if instead of the total group $Sp(4,\bf C)$
we consider its open subgroups $B_+=HN_+$ and $B_-=N_-H$ (this
means that the total symmetry is spontaneously broken and reduced up
to its open subgroups $B_+$ and $B_-$; by the way, the existence
of open subgroups was always some mystary which now clears little by
little) then we shall be able to integrate the equations on subgroups
$B_+$ and $B_-$.

      Clearly, the spontaneous breaking is not combined with the
unitarity using one self-adjoint space of representation.
But this phenomenon arises as a matter of course in the non-unitary
scheme using a dual pair of spaces $(\dot{\bbox{\cal F}},\bbox{\cal F})$,
$\dot{\bbox{\cal F}}\not=\bbox{\cal F}$ with a non-Hermitian form
$\langle\cdot,\cdot\rangle$. In this case the group variables may be
separated, so that the subgroup $B_+$ will be represented only (its
Lie algebra $b_+$ will be integrable) in one from the spaces, say in
$\bbox{\cal F}$, and $B_-$ (and $b_-$) will only in $\dot{\bbox{\cal F}}$
(their topology in $\bbox{\cal F}$ and $\dot{\bbox{\cal F}}$ is
selected such that the nilpotent algebra $\it n_-$ is not integrated
in $\bbox{\cal F}$ as $\it n_+$ in the case of $\dot{\bbox{\cal F}}$
\cite{10}). Here the Gaussian decomposition $N_-HN_+$ for the group
$Sp^{(*)}(4,\bf C)$ is selected as $p_\mu\in N_+$, $\dot p_\mu\in N_-$.
Thus the operators $p_\mu$ and $\dot p_\mu$ non-commuting with each
other generate flows in different spaces: respectively in $\bbox{\cal F}$
and in $\dot{\bbox{\cal F}}$ being dual to $\bbox{\cal F}$, so that
the equations of fluxes are written in the form
\begin{equation}
-i\frac{\partial}{\partial x_\mu}f(x)=p_\mu f(x),\quad
-i\frac{\partial}{\partial \dot x_\mu}\dot f(\dot x)=
\dot p_\mu \dot f(\dot x)
\label{eq10}
\end{equation}
where $f(x)\in \bbox{\cal F}$, $\dot f(\dot x)\in\dot{\bbox{\cal F}}$.
The infinite-component fields $f(x)$ and $\dot f(\dot x)$ existing
in a fibre ($x$ and $\dot x$ are coordinates in the fibre) are
transformed by infinite-dimensional representations of the group
$SU(2)$ with quarter-integer spins and are called as semispinor
ones \cite{10}. It is of vital importance that in the non-unitary
theory the Lorentz group $GL(2,{\bf C})\,\subset Sp^{(*)}(4,\bf C)$,
too, disintegrates in its Borel subgroups as
$N_\pm(GL(2,{\bf C}))\,\subset N_\pm(Sp^{(*)}(4,\bf C))$. In this
way the Lorentz symmetry of relativistic bi-Hamiltonian system
turns out to be spontaneously broken.

      The decomposition $(B_-,B_+)$ co-ordinated to the pair of
spaces $(\dot{\bbox{\cal F}},\bbox{\cal F})$ (in \cite{10} the
correspondence is called the polarization of dynamical system) is
responsible for the spontaneous '-symmetry (time boost) breaking.

      In the non-unitary theory the functions of two variables $x$ and
$\dot x$ exist only as sesquilinear non-Hermitian forms
$\langle\dot f(\dot x),f(x)\rangle$ or matrix elements being the
finite-component bilocal fields satisfying the condition
$$\left(\frac{\partial}{\partial x_\mu}\frac{\partial}
{\partial\dot x_\nu}-\frac{\partial}{\partial\dot x_\nu}
\frac{\partial}{\partial x_\mu}\right)\,
\langle\dot f(\dot x),f(x)\rangle\,=\,0\,.$$
In the theory such fields correspond to the
non-point-like fundamental particles. The manifold $M=\bf A_{3,1}$
will be reconstructed by them, and if the quantization procedure
applies to them we arrive at the theory without ultraviolet
divergences \cite{8}. Thus before using the second quantization
procedure it is necessary to carry out changing the theory
proposed here.

      For the present we have given the elementary introduction (from
authors' standpoint) to the class of ideas which would taken as a basis
for construction of a new theory of elementary particles. As seen
from the above, the proposed theory is related to the fundamental
break-up of usual notions (space-time continuum, unitarity axiom,
Hamiltonian systems) and their change by others (discontinuum,
non-unitary representations, non-standard dynamical systems).
The new theory is based on the notion of the relativistic
be-Hamiltonian system which was succeeded in constructing the
Lagrangian field system characterized by a certain mass spectrum
and interactions. The questions of reconstruction will be considered
elsewhere.

\end{document}